\documentclass[pra,aps,twocolumn,superscriptaddress]{revtex4}
\usepackage{epsf}
\usepackage{amsmath}
\usepackage{amssymb}

\usepackage{epstopdf}
\usepackage{graphicx}

\newtheorem{theorem}{Theorem}

\def \etal{\textit{et al.}}
\def \Tr{\mathrm{tr}}

\begin{document}

\title{Analytical progress on symmetric geometric discord: Measurement-based upper bounds}

\author{Adam Miranowicz}

\affiliation{Faculty of Physics, Adam Mickiewicz University,
PL-61-614 Pozna\'n, Poland}

\author{Pawe\l{} Horodecki}

\affiliation{Faculty of Applied Physics and Mathematics,
Technical University of Gda\'nsk, PL-80-952 Gda\'nsk, Poland}

\affiliation{National Quantum Information Centre of Gda\'nsk,
PL-81-824 Sopot, Poland}

\author{Ravindra W. Chhajlany}

\affiliation{Faculty of Physics, Adam Mickiewicz University,
PL-61-614 Pozna\'n, Poland}

\author{Jan Tuziemski}

\affiliation{Faculty of Applied Physics and Mathematics,
Technical University of Gda\'nsk, PL-80-952 Gda\'nsk, Poland}

\affiliation{National Quantum Information Centre of Gda\'nsk,
PL-81-824 Sopot, Poland}

\author{Jan Sperling}

\affiliation{Arbeitsgruppe Quantenoptik, Institut f\"ur
Physik, Universit\"at Rostock, D-18051 Rostock, Germany}

\begin{abstract}
Quantum correlations may be measured by means of the distance
of the state to the subclass of states $\Omega $ having well
defined classical properties. In particular,  a geometric
measure of asymmetric discord [Daki\'c et al., Phys. Rev.
Lett. \textbf{105}, 190502 (2010)] was recently defined as
the Hilbert-Schmidt distance of a given two-qubit state to
the closest classical-quantum (CQ) correlated state. We
analyze a geometric measure of symmetric discord defined as
the Hilbert-Schmidt distance of a given state to the closest
classical-classical (CC) correlated state. The optimal member
of $\Omega $ is just specially measured original state both
for the CQ and CC discords. This implies that this  measure
is equal to quantum deficit of post-measurement purity. We
discuss some general relations between the CC discords and
explain why an analytical formula for the CC discord,
contrary to the CQ discord, can hardly be found even for a
general two-qubit state. Instead of such exact formula, we
find simple analytical measurement-based upper bounds for the
CC discord which, as we show, are very efficient in the case
of two qubits and may serve as independent indicators of
two-party quantum correlations. In particular, we propose an
adaptive upper bound, which corresponds to the optimal states
induced by single-party measurements: optimal measurement on
one of the parties determines an optimal measurement on the
other party. We discuss how to refine the adaptive upper
bound by nonoptimal single-party measurements and by an
iterative procedure which usually rapidly converges to the CC
discord. We also raise the question of optimality of the
symmetric measurements realising the CC discord on symmetric
states, and give partial answer for the qubit case.
\end{abstract}



\date{\today}

\maketitle

\pagenumbering{arabic}

\section{Introduction}

Entanglement is a fundamental type of quantum correlation
that has come to be seen as an important resource in Quantum
Information (see, e.g., Ref.~\cite{Horodecki09}). However, quantum mechanics
supports other, distinct from entanglement, types of quantum
correlations in composite systems, such as the so-called
quantum discord~\cite{Ollivier01,Henderson01}, whose
characterization is the topic of much current research (see
the review~\cite{Modi12} and references therein). Quantum
discord is an information-theoretic measure of correlations
where quantum correlations are identified in terms of the
difference of two classically equivalent definitions of
mutual information~\cite{Ollivier01,Henderson01} in a
composite system.  A different possible perspective on
quantumness of correlations is captured in terms of quantum
deficit functions~\cite{Oppenheim02}, {\it i.e.}, differences
between certain properties of a state, before and after
classical type measurements are performed on it. One such
important property is the optimal thermodynamic work that can
be extracted from a state in scenarios of classical (local)
measurement complemented by zero-, one- and two-way classical
communication between measuring parties~\cite{Oppenheim02} (a
state is classical if the deficit is zero). While the two-way
scenario is rather involved, the zero- and one-way quantum
work deficits are simply equal to the so-called relative
entropy of quantumness~\cite{Horodecki05,Modi10} - the
minimal entropic ``distance'' measure to specific classes of
classical-type states.

Distance measures to sets of states with only classical
correlations are promising, and conceptually, simple ways of
identifying quantum correlations. Recently, {\it e.g.},
Daki\'c \etal~\cite{Dakic10} introduced a geometric measure
of discord of a state as its minimal Hilbert-Schmidt distance metric to the
set of states with null quantum discord (these states are
one-side classical, or so-called classical-quantum (CQ)
states of the form $\rho = \sum_{i} p_{i} P_{i} \otimes
\rho_{i}$, where $P_i$'s are orthogonal projections with rank
one and $\rho_{i}$'s are quantum states).

A natural, symmetric measure of quantum correlations can be
obtained by constraining to a set of fully classical states,
{\it i.e.}, classical-classical (CC) states which are
diagonal in some product basis~\cite{Shi11}. The
optimization process required in the evaluation of (general)
quantum correlation measures renders their calculation
challenging. Here, we shall build on an equivalence between
geometric measures of quantum discord and quantum deficits of
purity to provide tight and faithful upper bounds on the
symmetric geometric discord.

The paper is organized as follows. In Sec. II, we provide
some basic definitions and theorems for the discords in
relation to quantum deficit. In Sec. III, we present our main
result -- the measurement-based upper bounds on the CC
discord. In Sec. IV, we give explicitly formulas for the
upper bounds in the case of two qubits. In Sec. V, we present
an analytical comparison of the discords and upper bounds for
some classes of states. In Sec. VI, we present a few methods
with examples for optimization of the upper bounds. We
conclude in Sec. VII.

\begin{figure}
\includegraphics[scale=0.2]{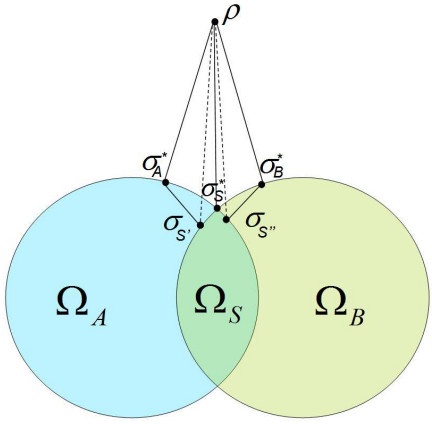}
\caption{(Color online) Venn-type diagram showing the sets of
the CC ($\Omega_{S}=\Omega_{AB}$), CQ ($\Omega_{A}$), and QC
($\Omega_B$) states together with the closest states
$\sigma^{*}_i$ ($i=S,A,B$) according to the CC ($D_S$), CQ
($D_A$), and QC ($D_B$) geometric discords, respectively.
States $\sigma_{S'}$ and $\sigma_{S''}$, which correspond to the
adaptive upper bound $D_S^{\rm (aub)}$, are the closest CC
states for $\sigma^*_{A}$ and $\sigma^*_{B}$, respectively.
This is an intuitive graph but, more precisely, the point
$\sigma^{*}_A$ ($\sigma^{*}_B$) should be on the line between
$\rho$ and $\sigma_{S'}$ ($\sigma_{S''}$). }
\end{figure}

\section{Background}

We start by recalling  the quantum zero-way and one-way work
deficit~\cite{Horodecki05}. Let the sets of states
$\Omega_{A}$, $\Omega_{B}$,  and $\Omega_{S}$ correspond to classical-quantum (CQ), quantum-classical (QC) and classical-classical (CC) states,
respectively (see Fig. 1). Note that the set
$\Omega_S\equiv\Omega_{AB}=\Omega_A\cap\Omega_B$  is
obviously in the intersection of the other two sets, and any
element of the intersection is in the set.
Let $\tilde{\cal M}_{X}$
correspond to all von Neumann's measurements that are
associated with the set $\Omega_{X}$ ($X=A,B,S$) in the
following  natural way,  $\tilde{\cal M}_{A}={\cal M}_{A}
\otimes I_{B}$, $\tilde{\cal M}_{B}= I_{A}\otimes {\cal
M}_{B}$ and $\tilde{\cal M}_{S}= {\cal M}_{A} \otimes {\cal
M}_{B}$, where ${\cal M}_{A}$, ${\cal M}_{B}$ are just local
von Neumann's measurements performed by Alice and Bob in some
orthonormal basis. We define the corresponding one-way ($X=A$
or $B$) and zero-way ($X=S$) quantum work deficits as:
\begin{equation}
\Delta_{X}(\rho)=\min_{\tilde{\cal M}_{X}} S[\tilde{\cal
M}_{X}(\rho)]- S(\rho), \label{Deficit}
\end{equation}
where $S(\cdot )$ is the von Neumann entropy. The relative
entropy of quantumness~\cite{Horodecki05,Modi10} is
\begin{equation}
D_{X}^{R}(\rho)=\min_{\sigma \in \Omega_{X}} S(\rho||\sigma).
\label{REQ}
\end{equation}

There is an observation (see Sec. VI.D in Ref. \cite{Horodecki05})
that links the above quantities.

{\it Observation 1 .--- For
any quantum state it holds that
$\Delta_{X}(\rho)=D_{X}^{R}(\rho)$. Furthermore
\begin{equation}
D_{X}^{R}(\rho)=\min_{\tilde{\cal M}_{X}} S(\rho||\tilde{\cal
M}_{X}(\rho)), \label{REQ1}
\end{equation}
which means that the optimal state $\sigma^{*}_{X}$,
saturating the minimum in Eq.~(\ref{REQ}), comes from the
optimal measurement in Eq.~(\ref{Deficit}) of the examined
state $\tilde{\cal M}_{X}^{*}(\rho)=\sigma^{*}_{X}$.} A proof
of Observation 1 is given in Appendix~\ref{Proofs}.

{\it Lemma 1.--- For any function $f$, any  Hermitian
operators $F$ and $G$, and any von Neumann's measurement operation
${\cal M}$ we have $\Tr[F f({\cal M}(G))]=\Tr[{\cal M}(F)
f({\cal M}(G))]$.} See Appendix~\ref{Proofs} for a proof of this Lemma.

{\it Geometric discord as a purity deficit.---} Geometric
measures of discord of a state are also similarly completely
determined by optimal measurements on it, as already noted by
Luo and Fu~\cite{Luo} and elucidated in the review by Modi
\etal~\cite{Modi12}. We formulate this property as follows:

{\it Observation 2.- Let $\sigma^{*}_{X} \in \Omega_{X}$ be
an optimal state saturating the minimum for quantum geometric
discord~\cite{Dakic10}:
\begin{equation}
D_{X}(\rho)=\min_{\sigma \in \Omega_{X}} ||\rho -
\sigma||^{2} \label{geometric}
\end{equation}
defined by the norm $||A||=\sqrt{\Tr(A^{\dagger}A)}$. Then it
is realised by some optimal measurement $\tilde{\cal
M}^{*}_{X}$ on $\rho$, {\it i.e.},
$\sigma^{*}_{X}=\tilde{\cal M}^{*}_{X}(\rho)$ and satisfies
the Pythagorean formula:
\begin{equation}
D_{X}(\rho)=||\rho -\sigma^{*}_{X} ||^{2}=||\rho||^{2} -
||\sigma^{*}_{X}||^{2}. \label{equivalence}
\end{equation}
Thus, $\tilde{\cal M}^{*}_{X}$  maximizes the
post-measurement purity  $\max_{\tilde{\cal M}_{X}} {\rm
\Tr}[(\tilde{\cal M}_{X}(\rho))^{2}]$ leading to the
alternative formula:
\begin{eqnarray}
D_{X}(\rho)&=&\min_{\tilde{\cal M}_{X}}||\rho -
\tilde{\cal M}_{X}(\rho)||^{2}   \nonumber \\
&=&\Tr(\rho^{2}) - \max_{\tilde{\cal M}_{X}} \Tr[(\tilde{\cal
M}_{X}(\rho))^{2}]. \label{equivalence1}
\end{eqnarray}}
See Appendix~\ref{Proofs} for a proof of this observation.
Note that choosing $X=A$ corresponds to one-side or
asymmetric (CQ) geometric discord~\cite{Dakic10}, while $X=S$
corresponds to the symmetric (CC) version~\cite{Shi11}. The
last form Eq.~(\ref{equivalence1}) for geometric discord
highlights an immediate analogy to the original deficit of
Eq.~(\ref{Deficit}) on replacing  the original von Neumann
entropy $S(\rho) \equiv S_{\alpha=1}(\rho)$ by the Tsallis
entropy
$S_{\alpha}(\rho)=-\frac{1}{\alpha-1}\Tr(\rho^{\alpha})$ (for
$\alpha=2$). We shall refer to the left-hand-side of Eq.
(\ref{equivalence1}) as a purity deficit, which is a special
case of entropy based deficits $\Delta_{X}^{\alpha,T}:=
\min_{ \tilde{\cal M}_{X}}S_{\alpha}[\tilde{\cal
M}_{X}(\rho)] - S_{\alpha}(\rho)$. For completeness, we
provide a proof of Observation 2 in Appendix~\ref{Proofs}
(see Ref.~\cite{Luo} and Secs. II.G and III.B.2 in
Ref.~\cite{Modi12} for alternate proofs).

{\it Simple consequences.---} Observation 2 leads us to:

{\it  Lemma 2.- For both the geometric discords $D_{X}$ and
quantum discords $D_{X}^{R}$ based on relative entropy, it
holds that: (i) The CQ and QC discords bound the CC discord
from below, $D_{S}(\rho) \ge \max[D_{A}(\rho),D_{B}(\rho)]$.
(ii) We have simple implications $D_{A}(\rho)=0 \Rightarrow
D_{S}(\rho)=D_{B}(\rho)$, and $D_{B}(\rho)=0 \Rightarrow
D_{S}(\rho)=D_{A}(\rho)$ with optimal measurement ${\cal
M}_{S}^{*}$ being a product of the measurement realising the
respective CQ or QC discord with the one that commutes with
the initially classical subsystem. (iii) $D_{S}(\rho)=0
\Leftrightarrow D_{A}(\rho)=D_{B}(\rho)=0$.}

See
Appendix~\ref{Proofs} for a proof of this Lemma.

\section{Measurement-based upper bounds on the CC discord}

We now turn to the main result of this paper. The explicit
calculation of the geometric CC discord is in general difficult as
it involves optimization over all measurements of the
required form given by Eq.~(\ref{equivalence1}). In
particular, the CC discord involves optimization over two
sides of the states and so involves twice as many parameters
as the CQ case. For the case of CQ-type discord, certain
lower bounds  have been found~\cite{Luo,Girolami12}.  On the other hand, we show here that the
measurement based formula Eq.~(\ref{equivalence1}) can be
fruitfully used to construct useful upper bounds on the CC
discord.

The conclusions of the present paragraph are valid for the
geometric and relative entropy discords and for  quantum
deficit based on any quantum entropy $S_{\alpha}$.

Let us
recall that one refers to a bound as: (i) {\it tight} if it
coincides with the bounded quantity on some non-trivial
subclass of states and  (ii) {\it faithful} iff it vanishes
on any state for which the bounded quantity vanishes.

\subsection{Nonadaptive upper bound}

An arbitrary measurement over two sides of the state is, by
definition, an upper bound on discords:
\begin{eqnarray}
D_{S}^{\sigma={\cal M}_{S}(\rho)}(\rho)&=& S_{\alpha}(\sigma)
- S_{\alpha}(\rho)\hspace{1cm}
\label{bound} \\
&=&S_{\alpha}[{\cal M}_{S}(\rho)]- S_{\alpha}(\rho) \geq
D_{S}(\rho),\nonumber
\end{eqnarray}
where $\alpha \in [0,\infty]$. For ease of notation, let
${\cal M}^{*}_{X,\rho}$ denote the optimal measurement
leading to the discord $D_{X}$ of state $\rho$. Product of
the two (CQ and QC) optimal measurements on the state $\rho$
leads to the first interesting bound, that we shall call the
{\it simple product  (or nonadaptive) bound}, for which the
measurement-induced state is:
\begin{equation}
\tilde{\sigma}=[{\cal M}^{*}_{A,\rho} \otimes {\cal
M}^{*}_{B,\rho}](\rho) \label{symmetric_bound}
\end{equation}
in Eq.~(\ref{bound}). This is one of the simplest kinds of
bounds  motivated by asking  how the CC and CQ discords (or
optimal measurements) are related. Indeed, we have already
noted in Lemma 2 that this type of bound trivially coincides
with the CC discord in the special case when one of the CQ
discords is null.

\subsection{Adaptive upper bound}

One  can further introduce refined bounds that are {\it
adaptive}, {\it i.e.},  measurement on one of the parties is
performed on the optimal state corresponding to the other
party, as below:
\begin{eqnarray}
&&\tilde{\sigma}'=[{\cal M}^{*}_{A,\rho} \otimes {\cal
M}^{*}_{B,{\cal M}^{*}_{A,\rho}(\rho)}](\rho),
\label{adapted1} \\
&& \tilde{\sigma}''= [{\cal M}^{*}_{A,{\cal
M}^{*}_{B,\rho}(\rho)} \otimes {\cal M}^{*}_{B,\rho}]
(\rho). \label{adapted2}
\end{eqnarray}
Note that part (ii) of  Lemma 2 immediately leads to the
following

{\it Fact 1.--- The bounds (\ref{bound}) based on
measurements, given by Eqs.~(\ref{symmetric_bound}),
(\ref{adapted1}), and~(\ref{adapted2}), are faithful, so they
may serve as independent indicators of two-side quantum
correlations.}

\subsection{Iterative procedure for the adaptive upper bound}

The adaptive form of Eqs.~(\ref{adapted1}) and (\ref{adapted2})
allows for an iterative procedure that may be helpful in
refining upper bound on the CC discord. Indeed, let $X$ and
$X'$ be two opposite subsystems [{\it i.e.}, $(X,X')=(A,B)$
or $(X,X')=(B,A)$]. Consider the following procedure: \emph{Step 1.}
Choose the initial subsystem $X=X_{0}$ (either $A$ or $B$),
and initial measurement ${\cal M}_{X}={\cal M}_{X_{0}}^{*}$.
\emph{Step 2}. Iterate the following steps: \emph{Step 2.1}. Given input
measurement ${\cal M}_{X}$ on $X$ calculate the output, {\it
i.e.}, optimal measurement ${\cal M}_{X',{\cal
M}_{X}(\rho)}^{*}$ on the second system $X'$. \emph{Step 2.2}. Put $X'$
in place of $X$ and the output ${\cal M}_{X',{\cal
M}_{X}(\rho)}^{*}$ as the input for Step 2.1, calculate its
output again. \emph{Step 2.3}. Calculate the bound on the discord, given
by Eq.~(\ref{bound}), with the help of the measurement ${\cal
M}_{S}$ being the tensor product of the input-output pairs of
measurements on $X$ and $X'$ presented in Steps 2.1 and
2.2, take the minimum of the two. \emph{Step 2.4}. Take the minimum of
the output of Step 2.3 of two subsequent rounds. \emph{Step 3}. Stop the
procedure if the outcome of Step 2.4 does not change.

\section{Two-qubit case revisited}

We now consider the case of the CC discord of two qubit states.
The standard Bloch representation of any two-qubit state is
\begin{eqnarray}
\rho  &=&  \frac{1}{4}(I\otimes
I+\vec{x}\cdot\vec{\sigma}\otimes
I+I\otimes\vec{y}\cdot\vec{\sigma}+\sum_{i,j=1}^{3}T_{ij}\sigma_{i}\otimes\sigma_{j})\nonumber
\\
&\equiv&  f(\vec{x},\vec{y},T)\equiv
f(|x\rangle,|y\rangle,T),
 \label{rho}
\end{eqnarray}
where $\vec{\sigma}=[\sigma_{1},\sigma_{2},\sigma_{3}]$ is a
vector of three Pauli matrices, $T$ is the correlation matrix
with elements
$T_{ij}=\mathrm{tr}[\rho(\sigma_{i}\otimes\sigma_{j})]$;
$\vec{x}=[x_{1},x_{2},x_{3}]^T\equiv|x\rangle$ and
$\vec{y}=[y_{1},y_{2},y_{3}]^T\equiv|y\rangle$ are the
(column) local Bloch vectors with components
$x_{i}=\mathrm{tr[}\rho(\sigma_{i}\otimes\openone)]$ and
$y_{i}=\mathrm{tr[}\rho(\openone\otimes\sigma_{i})]$.

\subsection{CC vs CQ discords}

We state the following simple

{\it Fact 2.--- Any  two-qubit state
$\rho=f(|x\rangle,|y\rangle,T)$  is mapped into
\begin{eqnarray}
\sigma_{(\hat{n})_{A}}(\rho)&=&f(|\hat{n}\rangle \langle
\hat{n} | x\rangle, |y\rangle,
|\hat{n}\rangle \langle \hat{n}|T), \\
\label{i} \sigma_{ (\hat{m})_{B}}(\rho)&=& f(|x\rangle,
|\hat{m}\rangle \langle \hat{m}|y\rangle,
T|\hat{m}\rangle \langle \hat{m}|), \\
\label{ii} \sigma_{ (\hat{n}, \hat{m})_{S}}(\rho) &=&
f(|\hat{n}\rangle \langle \hat{n} | x\rangle, |\hat{m}\rangle
\langle \hat{m} |y\rangle, |\hat{n}\rangle \langle \hat{n}
|T|\hat{m}\rangle \langle \hat{m}| ) \label{iii}\hspace{6mm}
\end{eqnarray}
by the measurement of  (i) $\hat{n}\vec{\sigma}$ on the left
qubits, (ii) $\hat{m}\vec{\sigma}$ on the right qubits, and
(iii) $\hat{n}\vec{\sigma}$ and $\hat{m}\vec{\sigma}$ on the
left and right qubits, respectively.}

This follows from  Lemma 1 and the fact that diagonal of
$\hat{n}\vec{\sigma}$ vanishes in the eigenbasis of any
$\hat{n}'\vec{\sigma}$ with $\hat{n}' \perp \hat{n}$.

Observation 2 and  Fact 2 directly  lead to the analytical
formula (see Ref.~\cite{Dakic10}) for the CQ discord $D_{A}$
as follows:
\begin{eqnarray}
D_{A}(\rho)&=& ||\rho||^{2} -
||\sigma^{*}_{A}||^{2}\nonumber \\ &=&\Tr(\rho^{2}) - \max_{\hat{k}}\{\Tr[(\sigma_{(\hat k)_{A}}(\rho))^{2}]\} \nonumber \\
&=& \frac{1}{4}\left(||\vec{x}||^{2} + ||T||^{2}-
\max_{\hat{k}}
[\langle \hat{k} (|x\rangle\langle x| + TT^{T} ) |\hat{k} \rangle ]\right)\ \nonumber \\
&=& \frac{1}{4}(||\vec{x}||^{2}+||T||^{2}-k_{x})=\frac{1}{4}(\Tr K_x-k_{x}) ,\label{Da}
\end{eqnarray}
where $\hat{k}_{x}$ is the largest eigenvalue of matrix
$K_{x}=|x\rangle\langle x| + TT^{\mathrm{T}}$. For clarity,
we also write\begin{eqnarray}
4||\rho||^{2}=&1+||\vec{x}||^{2}+||\vec{y}||^{2}+||T||^{2}\nonumber
\\ \equiv& 1+\langle x|x\rangle+\langle
y|y\rangle+{\rm tr}(TT^{T}).\label{norm_rho}
\end{eqnarray}
However  Observation 2 yields more, {\it viz.} the
eigenvector $|\hat{k}_{x}\rangle $ corresponding to the
eigenvalue $k_{x}$ defines the optimal measurement of party A
on $\rho$ producing the closest CQ state $\sigma$, which
({\it via} the Fact 2)  is
\begin{eqnarray}
\sigma_{{A}}^{*}= f(\langle
\hat{k}_{x}|x\rangle|\hat{k}_{x}\rangle,|y\rangle,|\hat{k}_{x}\rangle\langle
\hat{k}_{x}|T).\label{sigma_a}
\end{eqnarray}
Analogously, one obtains $D_{B}(\rho)=\frac{1}{4}(\Tr K_y-k_{y})=
\frac{1}{4}(||\vec{y}||^{2}+||T||^{2}-k_{y}),$ where $k_{y}$
is the largest eigenvalue of matrix $K_{y}=|y\rangle\langle
y|+T^{\mathrm{T}}T$ with the eigenvector
$|\hat{k}_{y}\rangle$. The closest QC state is
\begin{eqnarray}
\sigma_{{B}}^{*}= f(|x\rangle,\langle
\hat{k}_{y}|y\rangle|\hat{k}_{y}\rangle,T|\hat{k}_{y}\rangle\langle
\hat{k}_{y}|).\label{sigma_b}
\end{eqnarray}
Observation 2 also  delivers the two-qubit CC discord
\begin{eqnarray}
&&D_{S}(\rho)= ||\rho||^{2} - ||\sigma^{*}_{S}||^{2}
\label{Ds}
\end{eqnarray}
with the norm of
$\sigma^*_S=f(|x^*_S\rangle,|y^*_S\rangle,T^*_S)$  can be
given in terms of some functions minimized solely over unit
vectors $|\hat{x}_{S}\rangle$ (or, equivalently,
$|\hat{y}_{S}\rangle$) as given in Appendix~\ref{Ds_num}.
This is identical to the single Bloch-sphere optima  obtained
in Ref.~\cite{Shi11}.

\subsection{Quest for symmetry of the optimal measurement for
symmetric states}

There is a general question whether the states symmetric
under swapping subsystems always allow for a symmetric
optimal measurement in the formula for the CC discord $D_S$.
Here, we provide some partial results on this problem.
Namely, there is a practical observation:
\begin{theorem}
Consider the two-qubit symmetric states $\rho,$ i.e., the
ones satisfying $\rho_{AB}=\rho_{BA}$ or, equivalently,
\begin{eqnarray}
&&T=T^{T}, \\
&& |x\rangle=|y\rangle.
\end{eqnarray}
If the matrix $T$ satisfying either $T\geq 0$ or $(-T)\geq 0$
then the optimal CC state  $\sigma^*_{S}$ and the
corresponding measurement are symmetric, i.e., the optimal
measurement basis is defined by some
$|\hat{x}^*_{S}\rangle=|\hat{y}^*_{S}\rangle$.
\label{t-practical}
\end{theorem}
{\it Proof.---} Clearly, since $D_{S}(\rho)= \Tr(\rho^{2}) -
\max_{\hat{x}_{S},\hat{y}_{S}}{\rm Tr}[\sigma_{({\hat
x}_{S},{\hat y}_{S})_{AB}} (\rho)^{2}]$, we may write it in
the form
\begin{equation}
D_{S}(\rho)=\Tr(\rho^{2}) -\frac{1}{4}
\Big[1+\max_{\hat{x}_{S},\hat{y}_{S}} u(\hat{x}_{S},\hat{y}_{S})\Big],
\label{formula-s}
\end{equation}
where the function $u$ is defined as
\begin{equation}
u(\hat{x}_{S},\hat{y}_{S}) \equiv \langle
\hat{x}_{S}|T|\hat{y}_{S}\rangle^{2} + \langle
\hat{x}_{S}|x\rangle^{2} + \langle \hat{y}_{S}|y\rangle^{2}.
\end{equation}
Following Theorem 1, it is not difficult to see that, by the
symmetry of the initial state $\rho,$ one has
$|x\rangle=|y\rangle$. Now for $T>0$ (all eigenvalues
strictly positive) one defines the new  scalar product
$(x_S,y_S)_{T}= \langle \sqrt{T}x|\sqrt{T}y\rangle, $ which
defines also the norm $|| \vec x_S||_{T}= \sqrt{(x_S,x_S)}_{T}$. Now,
since $||\vec{x}_S - \vec{y}_S||^{2}_{T} \geq 0$, for any pair of unit
vectors $|\hat{x}_{S}\rangle$ and $|\hat{y}_{S}\rangle$ one
has $\frac{1}{2}[u(\hat{x}_{S},\hat{x}_{S}) +
u(\hat{y}_{S},\hat{y}_{S})] \geq u(\hat{x}_{S},\hat{y}_{S}),$
which means  that the maximum in Eq. (\ref{formula-s}) is
achieved by a symmetric pair
($|\hat{x}^*_{S}\rangle=|\hat{y}^*_{S}\rangle$). The proof
for $(-T)>0$  goes along the same lines.  For the cases when
$T\geq 0$ or $(-T)\geq 0,$ i.e., where zero eigenvalues are
allowed, the statement follows from the continuity argument
since here the argument realizing the maximum is continuous
in parameters of the state.

We conjecture that in the case of the symmetric two-qubit
states any minimum can be reached by symmetric measurement.
We have performed both analytical and numerical search and
found no counterexample to this hypothesis so far. However
for higher dimensions it may not be true since as we know
there are numerous properties that break there.

\subsection{Adaptive and nonadaptive upper bounds}

We now turn to the upper bound for the CC discord.  Using the
adaptively measured states, given by Eqs.~(\ref{adapted1})
and (\ref{adapted2}), we obtain the following upper bound
from Eq.~(\ref{bound}) for $\alpha=2$.

{\it Theorem.--- For an arbitrary two-qubit state the
adaptive upper bound $D^{\rm (aub)}_{S}(\rho)$ for the CC
discord can be given by:
\begin{eqnarray}
D_{S}^{{\rm (aub)}}(\rho)
=\min_{i=S',S''}||\rho-\sigma_{i}||^{2}=||\rho||^{2}-\max_{i}||\sigma_{i}||^{2},
\label{Ds_up}
\end{eqnarray}
where the CC states $\sigma_{S'}$ and $\sigma_{S''}$ are
\begin{eqnarray}
\sigma_{S'}&=&f(|x_{S'}\rangle,|y_{S'}\rangle,T_{S'}) =
\sigma_{( \hat{k}_{x}, \hat{l}_{y})_{S}}(\rho)
\label{sigma_s1}\\
&=&f(\langle \hat{k}_{x}|x\rangle|\hat{k}_{x}\rangle,\langle
\hat{l}_{y}|y\rangle|\hat{l}_{y}\rangle,|\hat{k}_{x}\rangle\langle
\hat{k}_{x}|T|\hat{l}_{y}\rangle\langle \hat{l}_{y}|),\nonumber\\
\sigma_{S''}&=&f(|x_{S''}\rangle,|y_{S''}\rangle,T_{S''})=\sigma_{
(\hat{l}_{x}, \hat{k}_{y})_{S}}(\rho)\label{sigma_s2}\\
&=&f(\langle \hat{l}_{x}|x\rangle|\hat{l}_{x}\rangle,\langle
\hat{k}_{y}|y\rangle|\hat{k}_{y}\rangle,|\hat{l}_{x}\rangle\langle
\hat{l}_{x}|T|\hat{k}_{y}\rangle\langle
\hat{k}_{y}|).\nonumber
\end{eqnarray}
where $|\hat{k}_{x}\rangle$, $|\hat{k}_{y}\rangle,$
$|\hat{l}_{x}\rangle$, and $|\hat{l}_{y}\rangle$ are the
eigenvectors corresponding to the maximum eigenvalue of}
\begin{eqnarray}
K_{x}&=&\vec{x}\vec{x}^{\mathrm{T}}+TT^{\mathrm{T}}\equiv|x\rangle\langle
x|+TT^{\mathrm{T}},\label{Kx} \\
K_{y}&=&|y\rangle\langle
y|+T^{T}T,\label{Ky} \\
L_{x}&=&|x\rangle\langle x|+T|\hat{k}_{y}\rangle\langle
\hat{k}_{y}|T^{\mathrm{T}},\label{Lx}\\
L_{y}&=&|y\rangle\langle y|+T^{T}|\hat{k}_{x}\rangle\langle
\hat{k}_{x}|T.\label{Ly}
\end{eqnarray}
respectively. Note that $\sigma_{S'}$  in general differs from
$\sigma_{S}^{*}$ used in Eq.~(\ref{Ds}). Explicitly, the
norms are given by
\begin{eqnarray}
||\sigma_{S'}||^{2}  = & \frac{1}{4}(1+||\vec{x}_{S'}||^{2}+||\vec{y}_{S'}||^{2}+||T_{S'}||^{2}) \nonumber \\
  = & \frac{1}{4}(1+\langle \hat{k}_{x}|x\rangle^{2}+\langle \hat{l}_{y}|y\rangle^{2}+\langle \hat{k}_{x}|T|\hat{l}_{y}\rangle^{2}), \label{norm_sigma_s1}
\\
||\sigma_{S''}||^{2}  = & \frac{1}{4}(1+||\vec{x}_{S''}||^{2}+||\vec{y}_{S''}||^{2}+||T_{S''}||^{2}) \nonumber \\
  = & \frac{1}{4}(1+\langle \hat{l}_{x}|x\rangle^{2}+\langle \hat{k}_{y}|y\rangle^{2}+\langle \hat{l}_{x}|T|\hat{k}_{y}\rangle^{2}), \label{norm_sigma_s2} \end{eqnarray}
Note that the measurement on direction $|\hat{l}_{x(y)}\rangle$
corresponds to the adaptive measurement $M^{\ast}_{A(B),
\rho}$, since the correlation matrix of the optimally
measured state on $A$ $(B)$ is according to
Eq.~(\ref{sigma_a}) [Eq.~(\ref{sigma_b})]  given by
$|\hat{k}_x \rangle\langle \hat{k}_x | T $  ($T|\hat{k}_y
\rangle\langle \hat{k}_y |  $).  Intuitively, the adaptive
upper bound $D_{S}^{{\rm (aub)}}(\rho)$ can also be found by
applying the following relation between the three discords
valid for an arbitrary two-qubit state: If $D_{A}(\rho)=0$
then $D_{S}(\rho)=D_{B}(\rho)$ and, analogously, if
$D_{B}(\rho)=0$ then $D_{S}(\rho)=D_{A}(\rho)$ as given by Lemma 2. The bound can
be constructed as follows (see Fig. 1):
\begin{equation}
D_{S}^{{\rm (aub)}}(\rho)=\min(D_{S'},D_{S''}),
\end{equation}
where
\begin{equation}
\begin{split}
 D_{S'}=||\rho-\sigma^{*}_{A}||^{2}+||\sigma^{*}_{A}-\sigma_{S'}||^{2}
 =||\rho||^{2}-||\sigma_{S'}||^{2},\hspace{2mm}\\
 D_{S''}=||\rho-\sigma^{*}_{B}||^{2}+||\sigma^{*}_{B}-\sigma_{S''}||^{2}
 =||\rho||^{2}-||\sigma_{S''}||^{2},
\end{split}
\end{equation}
and $\sigma_{S'}$ and $\sigma_{S''}$, given by
Eqs.~(\ref{sigma_s1}) and~(\ref{sigma_s2}), were calculated
from the repeated application of Eqs.~(\ref{sigma_a})
and~(\ref{sigma_b}). It is also worth noting that
\begin{equation}
 D_{S}^{{\rm  (aub)}}(\rho)=0 \Leftrightarrow
D_{S}(\rho)=0 \Leftrightarrow
D_{A}(\rho)=D_{B}(\rho)=0.\label{lemma4}
\end{equation}
So, in particular, it means that  $D_{S}^{{\rm (aub)}}(\rho)$
is nonzero iff $\rho$ is not a CC state, and thus it may
serve as an indicator of quantum  correlations itself.

The nonadaptive upper bound (i.e., product bound) for a two-qubit
state $\rho$ can be given by
\begin{equation}
D_{S}^{{\rm (nub)}}(\rho) =||\rho||^{2}-||\sigma_{S_0}||^{2},
\label{Dp_nub}
\end{equation}
where
\begin{eqnarray}
\sigma_{S_0}= f(\langle
\hat{k}_{x}|x\rangle|\hat{k}_{x}\rangle,\langle
\hat{k}_{y}|y\rangle|\hat{k}_{y}\rangle,|\hat{k}_{x}\rangle\langle
\hat{k}_{x}|T|\hat{k}_{y}\rangle\langle
\hat{k}_{y}|),\label{sigma_s0}
\end{eqnarray}
for which
\begin{equation}
||\sigma_{S_0}||^{2}=\frac{1}{4}(1+\langle
\hat{k}_{x}|x\rangle^{2}+\langle
\hat{k}_{y}|y\rangle^{2}+\langle
\hat{k}_{x}|T|\hat{k}_{y}\rangle^{2}). \label{norm_sigma_s0}
\end{equation}We have the following inequalities
\begin{equation}
\max(D_{A},D_{B})\le D_{S} \le D^{\rm (aub)}_{S} \le D^{\rm
(nub)}_{S}, \label{inequalities}
\end{equation}
where the last inequality can be immediately concluded by
comparing Eqs.~(\ref{norm_sigma_s1}) and
(\ref{norm_sigma_s2}) with~Eq.~(\ref{norm_sigma_s0}).

We note here that the adaptive bound, given by Eq.~(\ref{Ds_up}), is very
effective. Indeed, the largest gap to the exact value $\delta=
D^{\rm (aub)}_{S}(\rho)-D_{S}(\rho)$, observed by us
numerically, is just a few percent, and it is usually of the
order 10$^{-4}$ or 10$^{-5}$ for randomly generated rank-4
states. Interestingly, we have also observed that it is
exactly zero for almost all classes of states for which there
are known analytical expressions for $D_{S}$.

\section{Discords and upper bounds for some classes of states}

\subsection{Examples of simple relation between discords and their upper bounds}

Here, we present some examples of analytical calculation of
the CQ and CC discords and the adaptive upper bound to show
their relations.

{\em Example 1.---} For (a) pure states, (b) Bell diagonal
states, and also for (c) states with both marginals
vanishing, {\it i.e.}, $|x\rangle=|y\rangle=0$, it
holds\begin{equation} D_{A}=D_{B}=D_{S}=D^{\rm
(aub)}_{S}.\label{D_equal}
\end{equation}
For these states, $D_{S}$ can be easily found by showing
explicitly that the lower bound $D_{A}=D_{B}$ is equal to the
upper bound $D^{\rm (aub)}_{S}$.

{\em Example 2.---} For states with maximally  mixed single
marginal, {\it e.g.}, $|x\rangle=0$ (and analogously for $|y\rangle=0$), we have
\begin{eqnarray}
4||\sigma^{*}_{S}||^{2} & =&1+\underbrace{\langle
x|\hat{x}^{*}_{S}\rangle^{2}}_{=0}+\langle
y|\hat{y}^{*}_{S}\rangle^{2}+\langle\hat{x}^{*}_{S}|T|\hat{y}^{*}_{S}\rangle^{2}
 \nonumber \\&=&  1+\langle y|\hat{y}_{S}^{*}\rangle^{2}+\langle\hat{x}_{S}^{*}|\left[T|\hat{y}_{S}^{*}\rangle\langle\hat{y}_{S}^{*}|T^{T}\right]|\hat{x}_{S}^{*}\rangle.\end{eqnarray}
Since $|\hat{x}_{S}^{*}\rangle$ maximizes $||\sigma_{S}^{}||^{2}$
then it holds
\begin{equation}
|\hat{x}_{S}^{*}\rangle=\frac{T|\hat{y}_{S}^{*}\rangle}{\sqrt{\langle\hat{y}_{S}^{*}|T^{T}T|\hat{y}_{S}^{*}\rangle}}.
\end{equation}
Thus, we obtain
\begin{equation}
\begin{split}
4||\sigma_{S}^{*}||^{2}  = & 1+\langle y|\hat{y}_{S}^{*}\rangle^{2}+\left[\langle\hat{y}_{S}^{*}|T^{T}\left(\frac{T|\hat{y}_{S}^{*}\rangle}{\sqrt{\langle\hat{y}_{S}^{*}|T^{T}T|\hat{y}_{S}^{*}\rangle}}\right)\right]^{2}\\
  = & 1+\langle y|\hat{y}_{S}^{*}\rangle^{2}+\langle\hat{y}_{S}|T^{T}T|\hat{y}_{S}^{*}\rangle\\
  = & 1+\langle\hat{y}_{S}^{*}|\left(|y\rangle\langle y|+T^{T}T\right)|\hat{y}_{S}^{*}\rangle\\
  = & 1+\max{\rm [eig}\left(|y\rangle\langle y|+T^{T}T\right)],
\end{split}
\end{equation}
so finally\begin{equation} D_{S}=\frac{1}{4}\left\{\langle
y|y\rangle+||T||^{2}-\max \left[{\rm eig}(|y\rangle\langle
y|+T^{T}T)\right]\right\},
\end{equation}
which is equal to the QC discord $D_{B}$ and the adaptive
upper bound $D^{\rm (aub)}_{S}$, which follows from a simple
direct calculation. By performing analogous derivation for
$|y\rangle  = 0$,
 we conclude
that\begin{equation}
\begin{split}
 |x\rangle  = & 0\Rightarrow D_{A}\le D_{B}=D_{S}=D^{\rm (aub)}_{S},\\
 |y\rangle  = & 0\Rightarrow D_{B}\le D_{A}=D_{S}=D^{\rm (aub)}_{S}.
\end{split}
\end{equation}

\begin{figure}
\includegraphics[scale=0.45]{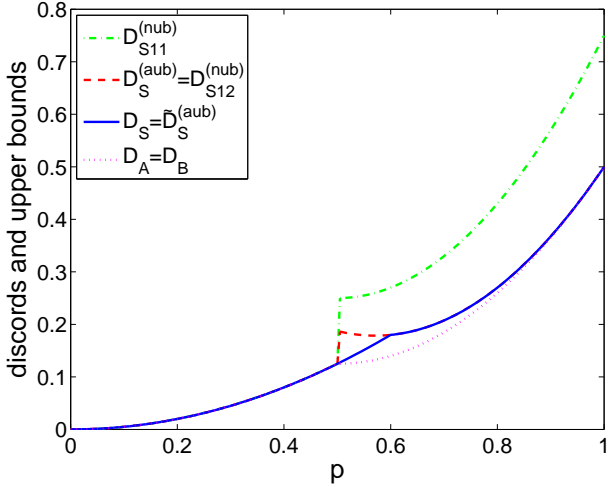}
\caption{(Color online) Geometric discords and their tight
upper bounds for the state $\rho(p,\phi=\pi/2)$, given by
Eq.~(\ref{h-state}), as a function of parameter $p$: The CC
discord, $D_{S}$ (blue solid), and CQ/QC discords,
$D_{A}=D_{B}$ (magenta dotted curve), together with the
adaptive upper bound, ${D}^{\rm (aub)}_{S}$ (red dashed), and
unoptimized nonadaptive upper bounds, $D_{S11}^{{\rm (nub)}}
=D_{S22}^{{\rm (nub)}}$ (green dot-dashed). We note that
$\tilde{D}^{\rm (aub)}_{S}=D_{S}$ and ${D}^{\rm
(aub)}_{S}=\tilde{D}^{\rm (nub)}_{S12}=\tilde{D}^{\rm
(nub)}_{S21}$ for any $p$. All the upper bounds, except
$\tilde{D}^{\rm (aub)}_{S}$, are discontinuous at $p=1/2$,
while the corresponding vertical lines are added for clarity
only.}
\end{figure}
\begin{figure}
\includegraphics[scale=0.45]{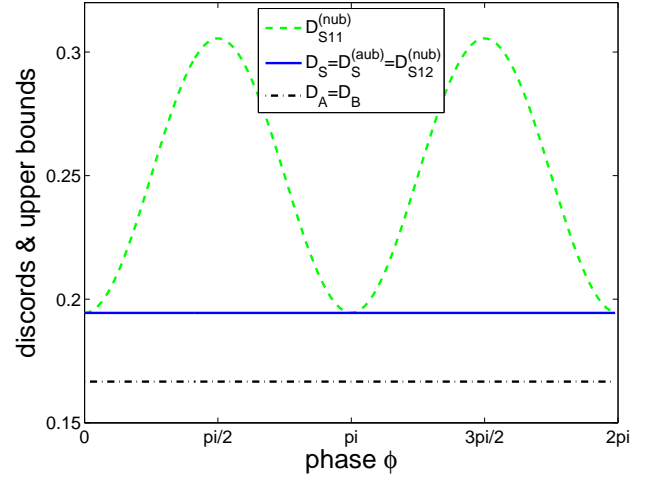}
\caption{(Color online) Same as in Fig. 2, but for the state
$\rho(p,\phi)$ as a function of phase $\phi$ for fixed
$p=2/3$. }
\end{figure}

\subsection{Example of nontrivial relation between discords and  their upper bounds}

Here, we give an example of nontrivial relation between CC,
CQ, and QC discords and their tight upper bounds as shown in
Figs. 2 and 3. Specifically, we will study mixtures of Bell's
state
$|\Psi_{\phi}\rangle=[|01\rangle+\exp(i\phi)|10\rangle]/\sqrt{2}$
and $|00\rangle$ (i.e., state separable and orthogonal to
$|\Psi_{\phi}\rangle$) as defined
by~\cite{Horodecki09,Miranowicz04}:
\begin{eqnarray}
\rho(p,\phi)=p|\Psi_{\phi}\rangle\langle\Psi_{\phi}|+(1-p)|00\rangle\langle00|\label{h-state}
\end{eqnarray}
for $0\le p\le1$. We find that the CC discord for these
states is given by\begin{eqnarray}
  \hspace{-5mm}D_{S}&=&\frac{1}{4}\min\left[2p^{2},7p^{2}-8p+3\right]
  \nonumber\\
  &=&\left\{ \begin{array}{c}
\frac{1}{2}p^{2}\hspace{22mm}{\rm if}\quad p\le\frac{3}{5},\\
\frac{1}{4}(7p^{2}-8p+3)\quad{\rm otherwise}.
\end{array}\right.\label{Ds_h}
\end{eqnarray}
By contrast, the CQ and QC discords are given by
\begin{eqnarray}
  D_{A}=D_{B}&=&\frac{1}{2}\min(p^{2},3p^{2}-3p+1)\nonumber\\
 &=&\left\{ \begin{array}{c}
\frac{1}{2}p^{2}\hspace{22mm}{\rm if}\quad p\le\frac{1}{2},\\
\frac{1}{2}(3p^{2}-3p+1)\quad{\rm otherwise.}
\end{array}\right.
\end{eqnarray}
Some details of the calculation of the discords are given in
Appendix~\ref{H-states}. Moreover, we find the adaptive upper
bound for the state $\rho(p,\phi)$ to be
\begin{eqnarray}
 & \hspace{-5mm}D_{S}^{{\rm (aub)}}\!=D_{S'}\!=\! D_{S''}\!=\left\{ \begin{array}{c}
\frac{1}{2}p^{2}\quad{\rm \hspace{18mm}if}\quad p\le\frac{1}{2},\\
\frac{1}{4}(7p^{2}-8p+3)\quad{\rm otherwise.}
\end{array}\right.\label{Ds_aub_h}
\end{eqnarray}
It is seen that $D_{S}^{{\rm (aub)}}=D_{S}$ for $0\le p\le 1/2$ and $3/5\le p\le 1$.

The nonadaptive and adaptive upper bounds for this state can
be optimized as will be described in the next section. All
these discords and upper bounds are shown in Fig. 2. In
particular, we observe discontinuity of the upper bounds at
$p=1/2$. We find that the upper bound $D_{S}^{{\rm (aub)}}$
(and $D_{S}^{{\rm (nub)}}$) has two different limits:
\begin{eqnarray}
\lim_{p\rightarrow1/2-}D_{S}^{{\rm (aub)}}  =  \frac{1}{8},\quad 
\lim_{p\rightarrow1/2+}D_{S}^{{\rm (aub)}}  = \frac{3}{16}.
\label{discontinuous}
\end{eqnarray}
By contrast, the asymmetric discords $D_{A}=D_{B}$, symmetric
discord $D_{S}$ and optimized upper bounds
$\tilde{D}_{S}^{{\rm (aub)}}$ (as discussed in the next
subsection) are continuous functions of any $p$. Anyway, none
of the discords has continuous first derivative in $p$.

\section{Improved upper bounds}

\subsection{Optimization over degenerate measurement outcomes}

If the maximal eigenvalues of operators $K_{x,y}$ and/or
$L_{x,y}$ are degenerate then the adaptive and nonadaptive
upper bounds can be optimized by taking  the minimum for the
eigenvectors corresponding these maximum eigenvalues. Here,
we will describe this method in brief and give an example
explaining Figs. 2 and 3.

First, it is worth recalling now a classic linear-algebraic
theorem stating that eigenvectors of degenerate matrices are
not necessarily orthogonal, but they can be made orthogonal
and complete, as in nondegenerate case, by applying
Gram-Schmidt's orthogonalization procedure. This is possible
by having additional freedom of replacing the eigenvectors
corresponding to a degenerate eigenvalue by their linear
combinations.

Let us denote  eigenvectors  $|\hat{k}_{x}^{(i)}\rangle$
($|\hat{k}_{y}^{(i)}\rangle$) corresponding to the same
maximum degenerate eigenvalue of operator $K_{x}$ ($K_{y}$),
given by Eq.~(\ref{Kx}) [Eq.~(\ref{Ky})]. Analogously, we
denote eigenvectors $|\hat{l}_{x}^{(ij)}\rangle$ and
$|\hat{l}_{y}^{(ij)}\rangle$ corresponding to the maximum
degenerate eigenvalues of operators:
\begin{eqnarray}
L_{x}^{(i)} & = & |x\rangle\langle
x|+T|\hat{k}_{y}^{(i)}\rangle \langle
\hat{k}_{y}^{(i)}|T^{\mathrm{T}}
,\label{Lx_i}\\
L_{y}^{(i)} & = & |y\rangle\langle
y|+T^{T}|\hat{k}_{x}^{(i)}\rangle\langle
\hat{k}_{x}^{(i)}|T,\label{Ly_i}
\end{eqnarray}
respectively. Thus, by applying these eigenvectors to Eqs.
(\ref{norm_sigma_s1}), (\ref{norm_sigma_s2}) and
(\ref{norm_sigma_s0}), one can obtain norms
$||\sigma_{S_0}^{(ij)}||^{2}$, $||\sigma_{S'}^{(ij)}||^{2}$
and $||\sigma_{S''}^{(ij)}||^{2}$ resulting in:
\begin{eqnarray}
{D}^{\rm
(aub)}_{Sij}&=&\min_{r=S',S''}(||\rho||^{2}-||\sigma_{r}^{(ij)}||^{2}),\\
{D}^{\rm
(nub)}_{Sij}&=&||\rho||^{2}-||\sigma_{S_{0}}^{(ij)}||^{2}.
\end{eqnarray}
Then, the optimized adaptive and nonadaptive upper bounds are
simply given by
\begin{eqnarray}
{D}^{\rm (aub)}_{S} =\min_{i,j}{D}^{\rm (aub)}_{Sij},\quad
{D}^{\rm (nub)}_{S} =\min_{i,j} {D}^{\rm
(nub)}_{Sij},\label{Daub_opt_deg}
\end{eqnarray}
respectively.

{\em Example.---}  Let us analyze again the state
$\rho(p,\phi)$, given by Eq. (\ref{h-state}). Operator $K_x$
is degenerate, as given by Eq.~(\ref{Kx_h-state}), so we can
choose $|\hat{k}_{x}^{(i)}\rangle=|i\rangle$. Simple
calculation shows that one can also choose
$|\hat{l}_{x}^{(ij)}\rangle=|j\rangle$ for $i,j=1,2$. We find
that the nonadaptive upper bound for $i=1,2$ is equal to
\begin{eqnarray}
 & D_{Sii}^{{\rm (nub)}}=\left\{ \begin{array}{c}
\frac{1}{2}p^{2}\hspace{31mm}{\rm if}\quad p\le\frac{1}{2},\\
\frac{1}{4}[p^{2}(\cos^{2}\phi)-8p+3]\quad{\rm otherwise.}
\end{array}\right. \label{nub1}
\end{eqnarray}
as shown by the green curves in Figs. 2 and 3. By contrast,
$D_{S12}^{{\rm (nub)}}=D_{S21}^{{\rm (nub)}}=D_{S}^{{\rm (aub)}}$
as  given by Eq. (\ref{Ds_aub_h}). So, finally,
\begin{eqnarray}
 & D_{S}^{{\rm (nub)}}=\min(D_{S11}^{{\rm (nub)}},D_{S12}^{{\rm (nub)}})= D_{S12}^{{\rm (nub)}}=D_{S}^{{\rm (aub)}}      \label{nub2}
\end{eqnarray}
as shown by the red curve in Fig. 2. Note that such
degenerate-value optimization for $D_{S}^{{\rm (aub)}}$ is
unnecessary for this state.

By analyzing our formulas and Fig. 2, we can observe that (1)
$D_{A}=D_{B}\neq D_{S}$ for $p\in(\frac{1}{2},1)$, (2)
$D_{S}^{{\rm (11)}}\neq D_{S}$ for $p\in(\frac{1}{2},1)$ if
$\phi\neq0$, and (3) $D_{S}^{{\rm (aub)}}=D_{S}^{{\rm
(nub)}}=D_{S11}^{{\rm (nub)}}(\phi=0)\neq D_{S}$ for
$p\in(\frac{1}{2},\frac{3}{5})$. We observe that the
unoptimized nonadaptive bound can be much greater than the
adaptive bound if $\phi\neq0$ and $\frac{1}{2}<p\le1$, thus
including the case for Bell's states ($p=1$). In Fig. 3, we
analyze the state $\rho(p,\phi)$ for $p=2/3$. We observe here
that (i) the symmetric discord (blue solid line) is equal to
the adaptive upper bound, $D_{S}={D}^{\rm (aub)}_{S}=7/36$,
(ii) the asymmetric discords (black dot-dashed line) are
$D_{A}=D_{B}=1/6$, (iii) the nonadaptive upper bound (red
dotted curve) depends on $\phi$ as follows $D_{S11}^{{\rm
(nub)}}=D_{S}+\sin^2(\phi)/9$. Finally, we conclude
$D_{S}=D^{\rm (aub)}_{S}= D_{S}^{{\rm (nub)}}= D_{S12}^{{\rm
(nub)}}\le D_{S11}^{{\rm (nub)}}\label{D_notequal1}$ for
$0\le p\le 1/2$ and $3/5\le p\le 1$. We see that the
nonadaptive bounds without optimization, on the other hand,
can fare rather badly as an estimator of the CC discord.

\subsection{Optimization by locally nonoptimal measurements}

Here, we suggest to optimize the adaptive upper bound by
locally nonoptimal measurements, i.e., to optimize over all
measurement outcomes corresponding to all (for $i,j=1,2,3$)
measurements of party A (B) on $\rho$ producing (usually not
the closest) state $\sigma^{(i)}_A$ ($\sigma^{(i)}_B$) and
then measurements of party B (A) on this  state producing the
state $\sigma^{(ij)}_{S'}$ ($\sigma^{(ij)}_{S''}$). Thus, we
describe the optimization of the adaptive upper bound over
all eigenvectors of $K_{m}$ and $L_{m}$ ($m=x,y$)
corresponding to all eigenvalues instead of taking only those
eigenvectors corresponding to the maximum eigenvalues of
these operators. This somehow counterintuitive method can be
in fact efficient for the adaptive upper bound since $L_{m}$
are constructed via eigenvectors of $K_{m}$. Clearly, the
method cannot improve the nonadaptive upper bound, as the
operators $L_{m}$ are not used there.

The optimized adaptive upper bound $\tilde{D}^{\rm
(aub)}_{S}$ can be defined in analogy to
Eq.~(\ref{Daub_opt_deg}) as follows:
\begin{equation}
\tilde {D}^{\rm (aub)}_{S} =\min_{i,j=1,2,3}D_{Sij}^{{ }}
=||\rho||^{2}-\max_{r=S',S''}\max_{i,j=1,2,3}||\sigma_{r}^{(ij)}||^{2}
.\label{Dup_opt}
\end{equation}
By contrast to Eq.~(\ref{Daub_opt_deg}), the optimalization
is over 2$\times$9  parameters for any state independent of
its degeneracy. It is convenient to form 3$\times$3 matrices
with elements $||\sigma_{r}^{(ij)}||^{2}$ as will be done in
the following.

{\em Example.---} Again we analyze the state, given by Eq.
(\ref{h-state}). For each of the three eigenvectors
$|\hat{k}_{m}^{(i)}\rangle$ of $K_{m}$ (for $m=x,y$), given
by Eq.~(\ref{Kx_h-state}), we find three orthogonal
eigenvectors $|\hat{l}_{m}^{(ij)}\rangle$, according to Eqs.
(\ref{Lx_i}) and (\ref{Ly_i}). Then, we can calculate
$||\sigma_{S'}^{(ij)}||^{2}=||\sigma_{S''}^{(ij)}||^{2}$ and
create, {\it e.g.}, the following matrices:
\begin{eqnarray}
[||\sigma_{S'}^{(ij)}||^{2}]=\left[\begin{array}{ccc}
\frac{1}{4} & A & C\\
\frac{1}{4} & A & C\\
C & C & B
\end{array}\right]\quad{\rm if}\quad p\le\frac{1}{2},
\end{eqnarray}
\begin{eqnarray}
[||\sigma_{S'}^{(ij)}||^{2}]=\left[\begin{array}{ccc}
C & C & B\\
\frac{1}{4} & C & A\\
\frac{1}{4} & C & A
\end{array}\right]\quad{\rm if}\quad p>\frac{1}{2},
\end{eqnarray}
where
 $A=(1+T_{11}^{2})/4=(1+p^{2})/4$,
 $B=(1+2x_{3}^{2}+T_{33}^{2})/4=[1+2(1-p)^{2}+(1-2p)^{2}]/4,$
 $C=(1+x_{3}^{2})/4=[1+(1-p)^{2}]/4$.
Any order of the eigenvectors (and, thus, the order of the
elements in the above matrices) can be applied. For
convenience, we ordered them here by the value of the
corresponding eigenvalues.
Then, we obtain
\begin{equation}
\tilde{D}^{\rm (aub)}_{S} =||\rho||^{2}-\max(A,B,C)=\left\{
\begin{array}{c}
||\rho||^{2}-B\quad{\rm if}\quad p\le\frac{3}{5},\nonumber\\
||\rho||^{2}-A\quad{\rm otherwise,}
\end{array}\right.
\end{equation}
where $||\rho||^{2}  =2p(p-1)+1$ (see Appendix C). Thus, we
conclude that the optimized upper bound is equal to the CC
discord for any $p\in[0,1]$:
\begin{equation}
\tilde{D}^{\rm (aub)}_{S}
=\frac{1}{4}\min\left[2p^{2},7p^{2}-8p+3\right]=D_{S}.\label{equal}
\end{equation}
in agreement with Eq. (\ref{Ds_h}). Note that $A=B$ for
$p=3/5$ and $A=C$ for $p=1/2$. So, $\tilde{D}^{\rm
(aub)}_{S}$ is continuous at $p=1/2$ contrary to
discontinuous $D_{S}^{{\rm (nub)}}$ and $D_{S}^{{\rm (aub)}}$
(compare broken and solid curves in Fig. 2).

In conclusion, for $\rho(p,\phi)$ with $1/2<p<3/5$ and
any $\phi$, we have the following inequalities
\begin{equation}
D_{A}=D_{B}< D_{S}=\tilde{D}^{\rm (aub)}_{S}<D^{\rm
(aub)}_{S}=D^{\rm (nub)}_{S}.\label{D_notequal2}
\end{equation}
This example demonstrates usefulness of the optimization
procedure by calculating the upper bounds for all possible
measurements rather than only for those measurements
corresponding to the maximum eigenvalues of $K_{m}$ and
$L_{m}$ ($m=x,y$).

\subsection{Iterative procedure for the adaptive upper bound\label{Iterative}}

Here, we describe in detail the iterative procedure described in Sec. III.C for the
adaptive upper bound $D^{\rm (aub)}$ and give some examples.
The $n$th iteration of the adaptive upper bound,
$D_{S}^{({\rm aub}\,n)}$, can be calculated as
\begin{eqnarray}
D_{S}^{({\rm
aub}\,n)}=||\rho||^{2}-\max(||\sigma_{S'}^{\{n\}}
||^{2},||\sigma_{S''}^{\{n\}}||^{2}),
\end{eqnarray}
where our old $D_{S}^{({\rm aub})}$ is just $D_{S}^{({\rm
aub}\,0)}$ and
\begin{eqnarray*}
||\sigma_{S'}^{\{n\}}||^{2}= & \frac{1}{4}(1+\langle \hat{k}_{x}^{\{n\}}|x\rangle^{2}+\langle \hat{l}_{y}^{\{n\}}|y\rangle^{2}+\langle \hat{k}_{x}^{\{n\}}|T|\hat{l}_{y}^{\{n\}}\rangle^{2}),\\
||\sigma_{S''}^{\{n\}}||^{2}= & \frac{1}{4}(1+\langle
\hat{l}_{x}^{\{n\}}|x\rangle^{2}+\langle
\hat{k}_{y}^{\{n\}}|y\rangle^{2}+\langle
\hat{l}_{x}^{\{n\}}|T|\hat{k}_{y}^{\{n\}}\rangle^{2}),
\end{eqnarray*}
where
$|\hat{k}_{x}^{\{n\}}\rangle=|\hat{l}_{x}^{\{n-1\}}\rangle$,
$|\hat{k}_{y}^{\{n\}}\rangle=|\hat{l}_{y}^{\{n-1\}}\rangle,$
while $|\hat{l}_{x}^{\{n\}}\rangle$,
$|\hat{l}_{y}^{\{n\}}\rangle$, $|\hat{k}_{x}^{\{0\}}\rangle$
and $|\hat{k}_{y}^{\{0\}}\rangle$ are the eigenvectors
corresponding to the maximum eigenvalues of
\begin{eqnarray}
L_{x}^{\{n\}} & = & |x\rangle\langle
x|+T|\hat{k}_{y}^{\{n\}}\rangle \langle
\hat{k}_{y}^{\{n\}}|T^{\mathrm{T}}
,\label{Lxx}\\
L_{y}^{\{n\}} & = & |y\rangle\langle
y|+T^{T}|\hat{k}_{x}^{\{n\}}\rangle\langle
\hat{k}_{x}^{\{n\}}|T,\\
K_{x}^{\{0\}} & \equiv & K_{x}= |x\rangle\langle
x|+TT^{\mathrm{T}},\\
K_{y}^{\{0\}}& \equiv &K_{y}=|y\rangle\langle
y|+T^{T}T,
\end{eqnarray}
respectively. For randomly generated rank-4 states (thus,
usually, with nondegenerate eigenvalues of $K_{x,y}^{\{0\}}$
and $L_{x,y}^{\{0\}}$),  the procedure is usually effective
as can be shown be calculating the difference
\[
\Delta_{n}\equiv D_{S}^{({\rm aub}\,n)}-D_{S}
\]
between the adaptive upper bound after the $n$th iteration
and the exact value of the CC discord.

Let us discuss just a few  examples:
\begin{table}
\caption{Examples of the application of the iteration
procedure for the adaptive upper bound $D_{S}^{({\rm aub})}$
for the states given by Eqs.~(\ref{state1}),~(\ref{state2}),
and ~(\ref{state3}) as described in Sec.~\ref{Iterative}. The
accuracy of the procedure is shown by the difference
$\Delta_{n}$ between the adaptive upper bound after the $n$th
iteration, $D_{S}^{({\rm aub}\,n)}$, and the exact value of
the CC discord, $D_{S}$.}
\begin{tabular}{l | l l l }
\hline\hline
iteration& \multicolumn{3}{c}{$\Delta_{n}=D_{S}^{({\rm aub}\,n)}-D_{S}$}\\
n &
state~(\ref{state1})\hspace{7mm} &
state~(\ref{state2})\hspace{7mm}  & state~(\ref{state3}) \\
\hline
$0$ & 8.85$\times10^{-4}$& 4.28$\times10^{-5}$ & 1.71$\times10^{-4}$  \\
$1$ & 0                  & 4.77$\times10^{-7}$ & 8.00$\times10^{-6}$  \\
$2$ & --                 & 5.53$\times10^{-9}$ & 3.90$\times10^{-7}$  \\
$3$ & --                 & 6.44$\times10^{-11}$& 1.92$\times10^{-8}$ \\
$4$ & --                 & $10^{-13}$          & 9.50$\times10^{-10}$\\
$5$ & --                 & $10^{-15}$          &
4.69$\times10^{-11}$\\ \hline \hline
\end{tabular}
\end{table}
{\em Example 1.---} Let us analyze state
$\rho=f(|x\rangle,|y\rangle,T)$ described by:
\begin{equation}
|x\rangle=|y\rangle=\frac{1}{4}[1,1,1]^{T},\quad
T=\frac{1}{4}{\rm diag}([1,-1,0]),\label{state1}
\end{equation}
First, we calculate the closest CQ state to be given in
Bloch's representation as
$\sigma^{*}_A=f(|x_A\rangle,|x\rangle,T_A)$, where
\begin{equation}
|x_{A}\rangle= t \left[
\begin{array}{c}
 1+\sqrt{3} \\
 1+\sqrt{3} \\
 2 \\
 \end{array}
\right],\quad T_A=\left[
\begin{array}{ccc}
 t & -t & 0 \\
 t & -t & 0 \\
 \frac{1}{8 \sqrt{3}} & -\frac{1}{8 \sqrt{3}} & 0
\end{array}
\right] \label{state1sigmaA}
\end{equation}
with $t=(3+\sqrt{3})/48.$ Analogously, the closest QC
state is $\sigma^{*}_B=f(|x\rangle,|x_A\rangle,T_A^T)$. Thus,
the CQ and QC discords are given by
$D_{A}=D_{B}=(3-\sqrt{3})/64=0.0198\cdots$. By contrast, the
closest CC state is much simpler as given by
$\sigma^{*}_S=f(|x\rangle,|x\rangle,Z)$, where $Z$ is the
zero matrix. Thus, the CC discord  is simply equal to
$D_{S}=1/32=0.031\cdots$. The nonadaptive upper bound is
$D_{S}^{\rm (nub)}=5(3-\sqrt{3})/192 =0.033\cdots$, which is
obtained as the Hilbert-Schmidt distance of $\rho$ to the CC
state $\sigma_{S_0}=f(|x_A\rangle,|x_A\rangle,Z)$, where
$|x_A\rangle$ is given in Eq.~(\ref{state1sigmaA}). The CC
states, defined by Eqs.~(\ref{sigma_s1}) and
(\ref{sigma_s2}), are equal to
$\sigma_{S'}=f(|x_A\rangle,|x\rangle,Z)$ and
$\sigma_{S''}=f(|x\rangle,|x_A\rangle,Z)$, respectively.
Thus, the (initial) adaptive upper bound is
$D_{S}^{\rm (aub)}\equiv D_{S}^{\rm (aub\,0)}=(21-5 \sqrt{3})/384=0.032\cdots$. Our
iteration procedure of the adaptive upper bounds converges to
$D_{S}$ already in the first iteration as $D_{S}^{({\rm
aub}\,1)}=1/32$ (see Table~I) since the CC state
$\sigma^{\{1\}}_{S'}=\sigma^{\{1\}}_{S''}=f(|x\rangle,|x\rangle,Z)=\sigma_{S}$.
Thus, for the analyzed state, we have the following
inequalities
\begin{equation}
D_{A}=D_{B}< D_{S}=D^{\rm (aub\,1)}_{S}< D^{\rm (aub)}_{S}
<D^{\rm (nub)}_{S}.\label{D_notequal3}
\end{equation}

{\em Example 2.---} Another state is given by the same
$|x\rangle=|y\rangle$ as in Eq. (\ref{state1}), but for
\begin{equation}
T=\frac{1}{4}{\rm diag}([1,1,0]).\label{state2}
\end{equation}
We find that the CQ/QC discords are
$D_{A}=D_{B}=(3-\sqrt{3})/64$ as in the former example, the
CC discord  is given by $D_{S}=0.02322\cdots,$ the
nonadaptive upper bound is $D_{S}^{\rm (nub)}= \left(28-11
\sqrt{3}\right)/384=0.02330\cdots$, and the adaptive upper bound is $D_{S}^{\rm (aub)}= \big[57-11
\sqrt{3}-\sqrt{6 (62+3 \sqrt{3})}\big]/768=0.02326\cdots$.
The iteration procedure converges to $D_{S}$ but not that
fast  as in the former example (see Table~I for details).

{\em Example 3.---} Now, let us analyze a state with
$|x\rangle\neq |y\rangle$ as defined by
\begin{equation}
|x\rangle=\frac{1}{6}[1,2,3]^{T},\;|y\rangle=\frac{6}{7}|x\rangle,\;
T=\frac{1}{8}{\rm diag}([1,2,3]).\label{state3}
\end{equation}
Analytical formulas for the discords and their upper bounds
are quite lengthy for this state, so we give only their
approximate numerical values. The QC discord is
$D_{B}\approx0.0259$, the CQ discord is $D_{A}\approx
0.0262$,  the CC discord is $D_{S}\approx0.0280$, the
nonadaptive upper bound is $D_{S}^{\rm (nub)}\approx0.0284$,
and the adaptive upper bound is $D_{S}^{\rm
(aub)}\approx0.0281$. The adaptive upper bounds $D_{S}^{({\rm
aub}\,n)}$ after the $n$th-iteration are shown in Table~I. In
conclusion, we have the following inequalities
\begin{equation}
D_{B}<D_{A}< D_{S}=D^{\rm (aub\,4)}_{S}-{\cal O}(10^{-10})<
D^{\rm (aub)}_{S} <D^{\rm (nub)}_{S}.\label{D_notequal4}
\end{equation}
All the above examples show how fast the iterations approach
the correct values of the CC discord. Now, we give a
counterexample:

{\em Example 4.---} The iteration procedure fails, e.g., for
the state, given by Eq. (\ref{h-state}) for $1/2<p<3/5$ (see
Fig. 2), as $\Delta_{n}=\Delta_{0}>0$ for $n=1,2,...$. In
general, this can be explained as follows:

\noindent \emph{Criterion.---} If, for a given two-qubit
state, the $n$th iteration of the adaptive upper bound $ D_{S}^{({\rm aub\,n})}$
(in particular for $n=0$) differs from the CC discord $D_{S}$, and
$|\hat{k}_{x}^{\{n\}}\rangle=|\hat{l}_{y}^{\{n\}}\rangle$ and
$|\hat{k}_{y}^{\{n\}}\rangle=|\hat{l}_{x}^{\{n\}}\rangle$
then the iteration procedure does not converge to $D_{S}$ as
$D_{S}^{({\rm aub,}n+k)}=D_{S}^{({\rm aub}\,n)}\neq D_{S}$ for
$k=1,2,...$.

Finally, we note that this iteration procedure can be improved
by replacing $D_{S}^{({\rm aub}\,n)}$ by the optimized
$\tilde D_{S}^{({\rm aub}\,n)}$ as described in the preceding
subsection.

\section{Conclusions}

We have shown that the geometric measures of quantum
correlations, i.e., the CC, CQ and QC discords, are equal to
the minimal purity deficit under specific von Neumann's
measurements compatible with the CC, CQ and QC classes of
states, respectively. This allowed us to quickly reproduce
known results in the case of qubits and also to give some
strong arguments that, the CC discord may not, in general, be
described analytically even for a two-qubit state. The best
general two-qubit formula, given by
Eqs.~(\ref{Bloch-sphere})--(\ref{fun_g}), still requires
minimalization over two variables. This is in contrast to the
CQ/QC discords for which analytical two-qubit  formulas are
available. Therefore, we focused on analytical approximations
of the CC discord. We proposed nonadaptive (i.e., simple
product) and adaptive upper bounds for the CC discord and
applied them for two-qubit states. We showed that they are
tight and faithful, so they can be used as independent tests
of nonclassical  quantum correlations. The adaptive upper
bound corresponds to an optimal measurement on one of the
parties conditioned an optimal measurement on the other
party. We also described a method of improving the adaptive
upper bound by nonoptimal single-party measurements. This
refined bound gives exact values of the CC discord for
(probably) all classes of states for which there are known
analytical expressions. For randomly generated states, the
bound usually differs from the CC discord by the order
10$^{-4}$ or 10$^{-5}$. Moreover, we described an iterative
procedure for the adaptive upper bound, which usually quickly
converges to the CC discord. We believe that this estimation
of the symmetric discord will play a role in analyzing the
cases when all the subsystems of a given quantum system
interact with the environment {\it on equal footing}. For
those cases it will probably be more adequate than asymmetric
discord that is based on system-apparatus picture.

\noindent {\bf Acknowledgments}. This work was supported by
the Polish National Science Centre under grants
2011/03/B/ST2/01903 and 2011/02/A/ST2/00305.

\appendix

\section{Proofs of observations and lemmas of Section II\label{Proofs}}

{\it Proof of Observation 1.---} Since ${\cal M}_{X}(\rho )
\subset \Omega _X$, by definition, the right-hand-side of
Eq.~(\ref{REQ}) is not greater than that of Eq. (\ref{REQ1}).
However, one can show that the opposite ordering of these
expressions can occur hence proving Observation 1. Indeed,
choose any measurement $\tilde{\cal M}_{X}$ commuting with
$\sigma^{*}_{X}$. Then the difference of Eqs.~(\ref{REQ}) and
(\ref{REQ1}) is equal to $S(\rho||\sigma^{*}_{X}) -
S(\rho||\tilde{\cal M}_{X}(\rho)) = S(\tilde{\cal
M}_{X}(\rho)||\sigma^{*}_{X})\geq 0$, where the first
equality is due to Lemma 1 below. Therefore $S(\tilde{\cal
M}_{X}(\rho)||\sigma^{*}_{X}) = 0$, {\it i.e.,}
$\sigma^{*}_{X} = \tilde{\cal M}_{X}(\rho),$ where the
measurement is the optimal one.  Lemma 1 further furnishes
the equivalence between the deficit $\Delta_{X}(\rho )$ and
$D_{X}^{R} (\rho )$ of Eq. (\ref{REQ1}).

{\it Proof of Lemma 1.---} Consider any von Neumann's
measurement ${\cal M}(\cdot)=\sum_i P_i (\cdot) P_i$ for
orthogonal projectors $\{ P_i \}$, $\sum_i P_{i}=I$. Since
any function of a Hermitian operator commutes with the
operator itself, one obtains $f({\cal M}(G))={\cal M} f
({\cal M}(G))$ and, consequently, $\Tr[ F f({\cal
M}(G))]=\Tr[ F {\cal M}(f({\cal M}(G))]= \Tr[ F \sum_i P_i
f({\cal M}(G))P_i]=\Tr[\sum_i P_i F P_i f ({\cal
M}(G))]=\Tr[{\cal M}(F)f({\cal M}(G))]$ for any Hermitian $F$
and $G$.

{\it Proof of Observation 2.---} To prove this Observation
for the symmetric discord $D_{S}$, we consider
\begin{eqnarray}
D_{S}(\rho)= \Tr(\rho^{2}) + \min_{\sigma \in \Omega_{S}} [
{\rm \Tr}(\sigma^{2}) - 2 \Tr(\rho\sigma)]
\label{def-general}
\end{eqnarray}
optimized over all the CC states $\sigma$ with eigenvectors
formed by two orthonormal bases ${\cal B}_{A} \otimes {\cal
B}_{B} := \{ |e_i\rangle |f_{j} \rangle \}$ and some
eigenvalues $\{ p_{ij}\} \equiv \vec{p}_{S}$. We may rewrite
Eq.~(\ref{def-general}) explicitly as
\begin{eqnarray}
D_{S}(\rho)=\Tr(\rho^{2}) -  \min_{ \vec{p}_{S}, {\cal B}_{A}
\otimes {\cal B}_{B} } [2 \Tr(\rho \sigma )-{\rm
\Tr}(\sigma^{2})]. \label{def-modified1}
\end{eqnarray}
Let  ${\cal B}^{*}_{A} \otimes {\cal B}^{*}_{B}$ be an
optimal basis defining naturally the von Neumann measurement
${\cal M}^{*}_{A} \otimes {\cal M}^{*}_{B} \equiv \tilde{\cal
M}_{S}^{*}$. The variational state defined in this basis of
course satisfies $\tilde{\cal M}_{S}^{*}(\sigma)=\sigma$ and
so (by Lemma 1) $\Tr( \rho \sigma )=\Tr[\tilde{\cal
M}_{S}^{*}(\rho) \sigma]$. Denoting by $\vec{q}_{S}$ the
diagonal of the state $\tilde{\cal M}^{*}_{S}(\rho)$ (which
is already of the CC type), we obtain $D_{S}(\rho)=
\Tr(\rho^{2})  -  \max_{\vec{p}_{S}} ( 2
\vec{q}_{S}\vec{p}_{S}  - \vec{p}_{S}^{\,2}),$ which yields
the optimal spectrum $\vec{p}^{\,*}_{S}=\vec{q}_{S}$. This
concludes the proof that the optimal state $\sigma^{*}$ in
Eq.~(\ref{def-general}) satisfies $\sigma^{*}=\tilde{\cal
M}_{S}^{*}(\rho)$. This combined with Lemma 1 implies both
Eqs.~(\ref{equivalence}) and (\ref{equivalence1}).

Consider now the asymmetric discord $D_A$, which is given by
\begin{eqnarray}
D_{A}(\rho)=\min_{\{ p_{i}, \sigma_{i} \}} \min_{{\cal
B}_{A}=\{e_{i}\}} [\Tr(\rho^{2}) - 2 \Tr(\rho \sigma ) +
\Tr(\sigma^{2}) ], \label{def-modified1-A}
\end{eqnarray}
where $\sigma=\sum_{i} p_{i} |e_{i}\rangle \langle e_{i}|
\otimes \sigma_{i} $. Let ${\cal B}^{*}_{A}=\{|
e_{i}^{*}\rangle\}$ be an optimal basis in Eq.
(\ref{def-modified1-A})  defining now the von Neumann
measurement ${\cal M}^{*}_{A}$ and the partially optimized
class of states $\sigma'=\sum_{i} p_{i} |e_{i}^{*}\rangle
\langle e_{i}^{*}| \otimes \sigma_{i}'$ which clearly
satisfies $[{\cal M}^{*}_{A} \otimes I_{B}](\sigma')=\sigma'
$. By Lemma 1,
\begin{equation}
D_{A}(\rho)= \Tr(\rho^{2}) + \min_{\{ p_{i},\sigma'_{i} \}}
\sum_i [ p_{i}^{2} \Tr( ( \sigma_{i} ' )^{2} )- 2 p_{i}q_{i}
\Tr( \sigma_{i}' \rho_{i}')], \label{def-modified2}
\end{equation}
where the parameters come from a new state
\begin{equation}
\rho' \equiv [{\cal M}^{*}_{A} \otimes I_{B}](\rho)=\sum_{i}
q_{i} |e_{i}^{*}\rangle \langle e_{i}^{*} | \otimes
\rho_{i}'. \label{rho-prime}
\end{equation}
For  all measurements ${\cal M}_{i}$ leaving $\rho_{i}'$-s
invariant we have $\Tr[(\sigma_{i} ' )^{2} ]= \Tr[{\cal
M}_{i}(\sigma_{i}')^{2}] + \delta_{i}$ (with $\delta_{i}\geq
0$), since von Neumann's measurements do not increase purity.
Using Lemma 1 again, we therefore have
\begin{eqnarray} &&\min_{\{
p_{i},\sigma'_{i} \}}
\sum_i [ p_{i}^{2} \Tr( ( \sigma_{i} ' )^{2} )  - 2 p_{i}q_{i} \Tr( \sigma_{i}' \rho_{i}')]   \nonumber \\
&&=\min_{\{ p_{i},\sigma'_{i} \}}
\sum_i [ p_{i}^{2} \Tr(  {\cal M}_{i}(\sigma_{i}')^{2}) + \delta_{i}  - 2 p_{i}q_{i} \Tr( {\cal M}_{i}(\sigma_{i}') \rho_{i}')]    \nonumber \\
&&\geq \min_{\{ p_{i},\sigma'_{i} \}} \sum_i [ p_{i}^{2}
\Tr({\cal M}_{i}(\sigma_{i}')^{2})   - 2 p_{i}q_{i} \Tr({\cal M}_{i}(\sigma_{i}') \rho_{i}')]   \nonumber \\
&& = \min_{\{ p_{i},\tilde{\sigma}'_{i}={\cal
M}_{i}(\tilde{\sigma}'_{i}) \}} \sum_i [ p_{i}^{2}
\Tr((\tilde{\sigma}_{i}')^{2})  - 2 p_{i}q_{i} \Tr(
\tilde{\sigma}_{i}' \rho_{i}')]
\nonumber \\
&&= \min_{\tilde{\sigma}'}[\Tr((\tilde{\sigma}')^{2}) -
2\Tr(\tilde{\sigma}'\rho')], \label{def-modified3-A}
\end{eqnarray}
where
\begin{equation}
\tilde{\sigma}'=\sum_{i} p_{i} |e_{i}^{*} \rangle \langle
e_{i}^{*}| \otimes {\cal M}_{i}(\sigma_{i}').
\label{partially-optimal-A}
\end{equation}
Since $\tilde{\sigma}'$ and $\rho'$  commute having product
eigenvectors (which, however, {\it do not form}  a product of
the two eigenbases in general), we are left only with the
final problem of  finding optimal eigenvalues of
$\tilde{\sigma}'$. In analogy to the solution of Eq.
(\ref{def-modified1}), one concludes immediately that the
optimal spectrum is the same as that of $\rho'$. Thus, the
optimal CQ state must be equal to $\rho'$, which is just the
original $\rho$ subjected to some specific measurement ${\cal
M}_{A}^{*} \otimes I_{B}$. This concludes the proof for the
CQ-type discord and thus, finally, the proof of Observation
2.

{\it Proof of Lemma 2.---} The property (i) is immediate
since the set $\Omega_{S}$ of all the CC correlated states is
a subset of the sets $\Omega_{A}$ and $\Omega_{B}$ of the
CQ/QC correlated states as shown intuitively in Fig. 1.
Property (ii) follows from the fact that for $D_{A}(\rho)=0$
the optimal measurement ${\cal M}^{*}_{B}$ providing $D_{B}$
already reduces the state $\rho$ to a CC state. This means
that combining it with the measurement ${\cal M}_{A}$
commuting with the left reduction of the state yields the
same value $D_{B}$. This value based on product measurement,
corresponds by definition Eq.~(\ref{geometric}) to some upper
bound of the CC discord $D_{S}$. On the other hand, $D_{B}$
is by (i) a lower bound on the CC discord. Thus
$D_{S}=D_{B}$. Property (iii) is implied by (ii).

\section{Numerical calculation of the CC discord \label{Ds_num}}

For completeness, we give an explicit formula for numerical
calculation of the CC discord for an arbitrary two-qubit
state $\rho$ as follows
\begin{equation}
D_{S}(\rho)=||\rho||^{2}
-\max_{\theta,\phi}||\sigma_{S}(\theta,\phi)||^{2}=||\rho||^{2}
-||\sigma^{*}_{S}||^{2},\!
\label{Bloch-sphere}
\end{equation}
where $\sigma_{S}$ [and thus
$\sigma^*_S=f(|x^*_S\rangle,|y^*_S\rangle,T^*_S)$] can be
expressed solely in terms of the versor
$$|\hat{x}_{S}\rangle=\frac{|x_{S}\rangle}{\sqrt{\langle
x_{S}|x_{S}\rangle}}=[\sin\theta\cos\phi,\sin\theta\sin\phi,\cos\theta]^{T},$$
as follows
\begin{eqnarray}
4||\sigma_S^*||^{2} &= &1+
 \langle{x}^*_{S}|x^*_{S}\rangle+\langle {y}^*_{S}|y^*_{S}\rangle
 +\langle{x}^*_{S}|T^*_{S}|{y}_{S}^*\rangle \nonumber\\
&= &1+
 \langle{x}^*_{S}|x\rangle+\langle {y}^*_{S}|y\rangle
 +\langle{x}^*_{S}|T|{y}_{S}^*\rangle \nonumber\\
&= &1+
 \max_{\hat{x}_{S},\hat{y}_{S}}\left(\langle\hat{x}_{S}|x\rangle^{2}+\langle _{}\hat{y}_{S}|y\rangle^{2}+\langle\hat{x}_{S}|T|\hat{y}_{S}\rangle^{2}\right) 
\nonumber
\\ &=& 1+ \max_{\hat{x}_{S}}
[ \lambda_{y}(\hat{x}_{S}) + \langle
\hat{x}_{S}|x\rangle^{2}] \nonumber
\\ &=& 1+ \max_{\hat{y}_{S}} [\lambda_{x}(\hat{y}_{S}) + \langle
\hat{y}_{S}|y\rangle^{2}],\label{norm_sigma}
\end{eqnarray}
where
the quantity $\lambda_{y}(\hat{x}_{S})$
[$\lambda_{x}(\hat{y}_{S})$] is the maximal
eigenvalue of the rank-two matrix $T|\hat{y}_{S}\rangle
\langle \hat{y}_{S}|T^{T} + |x\rangle \langle x|$
($T^{T}|\hat{x}_{S}\rangle \langle \hat{x}_{S}|T + |y\rangle
\langle y| $). So, explicitly,
\begin{eqnarray}
  \lambda_{y}(\hat{x}_{S})=h_{+}+\!\sqrt{\langle\hat{x}_{S}|T|y\rangle^{2}+h_{-}^{2}},
 \label{Pawel1}
\end{eqnarray}
and\begin{eqnarray}
  \lambda_{x}(\hat{y}_{S}) =g_{+}+\!\sqrt{\langle x|T|\hat{y}_{S}\rangle^{2}+g_{-}^{2}}, \label{Pawel2}
\end{eqnarray}
where
\begin{eqnarray}
h_{\pm}=&\frac{1}{2}(\langle
y|y\rangle\pm\langle\hat{x}_{S}|TT^{T}|\hat{x}_{S}\rangle),\label{fun_h}
\\
g_{\pm}=&\frac{1}{2}(\langle
x|x\rangle\pm\langle\hat{y}_{S}|T^{T}T|\hat{y}_{S}\rangle),\label{fun_g}
\end{eqnarray}
in agreement with Ref.~\cite{Shi11}.

\section{Calculation of discords for mixtures of $|00\rangle$
and Bell's states \label{H-states}}

Here, we give more details of our calculation of the CC and
CQ/QC discords for the states $\rho(p,\phi)$ defined by
Eq.~(\ref{h-state}).

The correlation matrix $T$ of Bloch's representation for the
state $\rho(p,\phi)$ reads as
\begin{equation}
T=\left[\begin{array}{ccc}
p\cos\phi & -p\sin\phi & 0\\
p\sin\phi & p\cos\phi & 0\\
0 & 0 & 1-2p
\end{array}\right],
\label{T_h-state}
\end{equation}
and the local Bloch's vectors are
$|x\rangle=|y\rangle=[0,0,1-p]^T$. First, we note that
\begin{equation}
K_{x}=K_{y}=|x\rangle\langle
x|+TT^{T}=\left[\begin{array}{ccc}
p^{2} & 0 & 0\\
0 & p^{2} & 0\\
0 & 0 & q
\end{array}\right],\label{Kx_h-state}
\end{equation}
where $q=(1-2p)^{2}+(1-p)^{2}$. Since $p^{2}\le q$ is
fulfilled if $p\in[0,\frac{1}{2}]$, so we have to analyze two
solutions for $p\le\frac{1}{2}$ and $p>\frac{1}{2}$ . Thus,
the CQ and QC discords are
\begin{equation}
\begin{split}
D_{A}  =D_{B}= & \frac{1}{4}\left[\Tr(K_{x})-\max {\rm eig}(K_{x})\right]\\
 =& \frac{1}{4}\left[7p^{2}-6p+2-\max(q,p^{2})\right]\\
 = & \frac{1}{2}\min(p^{2},3p^{2}-3p+1).
\end{split}
\end{equation}
We can also calculate the CC discord as follows. The norm
$||\sigma_{S}||^{2}$ is given by\begin{equation}
\begin{split}
4||\sigma_{S}||^{2}  -1= & \max_{\hat x_{S},\hat
y_{S}} \left(\langle\hat{x}_{S}|x\rangle^{2}+ \langle
\hat{y}_{S}|y\rangle^{2}+
\langle\hat{x}_{S}|T|\hat{y}_{S}\rangle^{2}\right)\\
  =& \max(\langle1|x\rangle^{2}+\langle1|y\rangle^{2}
+\langle1|T|1\rangle^{2},\\
  &\hspace{10mm}\langle3|x\rangle^{2}+\langle3|y\rangle^{2}
  +\langle3|T|3\rangle^{2})\\    = &
  \max\left[p^{2},2(1-p)^{2}+(1-2p)^{2}\right].
\end{split}
\end{equation}
Moreover, $||\rho||^{2}  =2p(p-1)+1,$ since $\langle
x|x\rangle=\langle y|y\rangle=(1-p)^{2}$ and
$||T||^{2}=(1-2p)^{2}+2p^{2}$. Thus, finally, we obtain
$D_{S}=||\rho||^{2}-||\sigma_{S}||^{2}$ given by Eq.~(\ref{Ds_h}). The geometric discords
$D_{A}=D_{B}$ and $D_{S}$ for this state are plotted in
Figs.~2 and 3.

\end{document}